\documentclass[aps,showpacs,pre,floatfix,twocolumn]{revtex4}
\usepackage{comment,psfig,amssymb,amsmath}
\def \ER{Erd\H{o}s-Renyi}

\newcommand{\Def}{\overset{\hbox{\it \small def}}{=}\ }

\newcommand{\Eqn}[1]{Eq.~(\ref{#1})}     
\newcommand{\Eqns}[1]{Eqns.~(\ref{#1})}     
\newcommand{\Sec}[1]{Section~\ref{#1}}     
     
\newcommand{\Fig}[1]{Fig.~\ref{#1}}     
\newcommand{\Figs}[1]{Figs.~\ref{#1}}

\newcommand{\peff}{\mathcal{P}}

\begin{document}
\title{Two rigidity percolation transitions on binary Bethe networks
  and intermediate phase in glass.}
%\title{Rigidity Percolation with two types of sites}
\author{Cristian  F.~Moukarzel
%~\footnote{email address: cristian@mda.cinvestav.mx}
}
\affiliation{Depto.\ de F\'\i sica Aplicada, CINVESTAV del IPN,\\
  Av.~Tecnol\'ogico Km 6, 97310 M\'erida, Yucat\'an, M\'exico } 
\date{\today}
\begin{abstract}
  Rigidity Percolation is studied analytically on randomly bonded
  networks with two types of nodes, respectively with coordination
  numbers $z_1$ and $z_2$, and with $g_1$ and $g_2$ degrees of freedom
  each. For certain cases that model chalcogenide glass networks, two
  transitions, both of first order, are found, with the first
  transition usually rather weak. The ensuing intermediate pase,
  although is not isostatic in its entirety, has very low
  self-stress. Our results suggest a possible mechanism for the
  appearance of intermediate phases in glass, that does not depend on
  a self-organization principle. 
\end{abstract}
%\pacs{}
\maketitle
\section{Introduction}
Structural (or Graph) Rigidity
(SR)~\cite{ARROG78,ARROG79,CSR79,LYOGR82,WIRP84} is the field of
applied mathematics that deals with the conditions under which a
structure, representable by a graph, can sustain mechanical loads. The
rudiments of SR were established by J.~C.~Maxwell in a classical
article~\cite{MCC64} that considered the balance of positional degrees
of freedom (at the nodes of the graph) and mechanical constraints
(edges in the graph). Rigidity, in the context or SR, is the vectorial
generalization of connectedness. A graph is connected if it can
transmit a current, while it is rigid if it can transmit a
force. Connectivity is a particular case of Rigidity, where all nodes
have only one degree of freedom (dof)~\cite{MDCOR99}.

Among the many applications of
SR~\cite{FSPOE84,WACB84,MDSBA95,TDRTA99,MGMI99,RHKPUR02,BJAFG03}, the
physics of Glass is a notable example~\cite{TDRTA99,MTCT11}, to which
the most important applications of the present work pertain.
Phillips~\cite{PTOC79} pioneered the field of SR in Glass, by using
constraint counting to explain the observed connection between
glass-forming ability and chemical composition in chalcogenides.
Thorpe~\cite{TCDI83} then discussed Phillips' ideas in the context of
Rigidity Percolation (RP). These early works established the bases for
the so-called Phillips-Thorpe vector percolation constraint
theory~\cite{PTCTV85} of chalcogenide glass, which has in later years
received ample support from experiments~\cite{SVCMST90,BECVTI90}. In
the original version of this theory, constraints are assumed
infinitely rigid, as appropriate for $T=0$. However, temperature
effects, e.g.~the fact that links can be broken if the available
thermal energy is large enough, can be included in several
ways~\cite{NELA05,GMCDO09}. Within a simple model, temperature effects
amount to random link dilution. The models discussed in this work thus
have two external parameters: chemical composition, i.e.~the relative
amount of each atomic species, and link density, loosely representing
temperature effects.

The original theoretical description of the link between rigidity and
glass-forming ability states that, as the chemical composition of a
chalcogenide glass is varied, rigidity percolates at a given
composition $x_c$, where the system goes from a flexible network to a
stressed rigid one~\cite{TCDI83,PTCTV85}. This critical concentration
is, ignoring deviations from Maxwell-Counting, determined by the
condition of \emph{global constraint balance} (GCB), i.e.~the
composition at which the total number of spatial degrees of freedom
equals the total number of mechanical constraints.

A more stringent condition, which is not implied by
GCB~\footnote{Despite wrong claims in numerous recent
  articles, where GCB=isostaticity is explicitely stated.}, is that of
\emph{isostaticity}, or \emph{local constraint balance} (LCB). The
concept of isostaticity was not part of the early picture for RP in
glass, which only discussed GCB.  Isostaticity, as a spontaneously
organized property of a natural system was first found in sphere
packings~\cite{MIPT98,MGMI99}, which are a model for metallic
glass. Isostaticity, or LCB, means that not only there is GCB, but
that no constraint is redundant. In other words, that constraints are
in such way distributed, that none of them is 'wasted' in an already
rigid region, and, therefore, that no degree of freedom is left
uncanceled in the system. This delicate balance is \emph{never}
satisfied by systems undergoing random RP, no matter what the density
of constraints is. Only those systems satisfy LCB, for which a
mechanism for constraint repulsion is at work, that disallows
redundant constraints. For sphere
packings~\cite{MIPT98,MGMI99,MRFI05}, this mechanism stems form the
fact that contact forces must all be compressive, which in turn
forbids \emph{cycles} (overconstrained subgraphs).  Thorpe and
coworkers~\cite{TJCSIN00,TCJNIN01,CTSAR01} later proposed an
energy-minimization mechanism for constraint repulsion, which appears
to be relevant in chalcogenide glasses. Their model predicts the
existence of two phase transitions, so it has three phases: floppy,
unstressed-rigid, and stressed-rigid. The intermediate phase is
composed of varying fractions of an isostatic-rigid phase, with floppy
pockets that disappear gradually towards the second transition, as
constraints are added to the system.  The existence of an intermediate
phase in chalcogenides, that appears to be nearly stress-free, had
been previously detected by
Boolchand~\cite{SBBTRW99,SBBSTI00,WBMGSR00,BFBRTI01} experimentally.

Thorpe and coworkers' self-organized
model~\cite{TJCSIN00,TCJNIN01,CTSAR01} explains the
appearance of the intermediate phase on the grounds that redundant
constraints introduce self-stress, and, therefore, increase the
elastic energy in the system. Therefore, when the glass is formed,
redundant constraints would be avoided in order to minimize the
energy. However, this line of reasoning might have to be revised in
the presence of strong enough external loads acting on the forming
glass, as recent work has shown that randomly disordered isostatic
networks have zero elastic modulus~\cite{MECI12}. Now, the
\emph{total} elastic energy of a rigid network has two parts: one is
due to self-stress, and this is effectively minimized for nearly
isostatic networks.  The second contribution to the elastic energy
comes from load-induced stress, and has the form \hbox{$E_{load} \sim
  L^2/B$}, where $L$ is an external load (e.g.~isotropic compression),
and $B$ is the elastic modulus that is relevant for that load
(e.g.~bulk modulus). This contribution is therefore \emph{maximized}
for an isostatic network, according to a recent
discussion~\cite{MECI12}.  Therefore, if the external load is large
enough, the total elastic energy might actually be smaller for an
overconstrained network than for an isostatic one, under the same set
of external conditions. This observation might suggest a possible
experimental test of Thorpe et al mechanism for isostaticity, as glass
formed under strong enough compression should depart more from
isostaticity, i.e.~have more self-stress, than glass formed under
load-free conditions. Bear in mind, however, that the scale of loads
needed to observe this effect in experiments might be too large to be
of any relevance.

On a different note, and despite the physical appeal of Thorpe and
coworkers ideas to explain the genesis of the intermediate phase, it
can be seen that, in fact, no self-organization principle or
constraint-repulsion mechanism is needed in order to obtain two RP
phase transitions and, therefore, an intermediate phase, in a
mean-field model of binary chalcogenide networks.  The main focus of
this work is to study mean-field RP on Bethe networks made from a
mixture of two types of sites with different coordination numbers,
i.e.~a toy model for a binary glass.  We will analytically study such
``mixed'' Bethe networks with arbitrary amounts of \emph{random}
bond-dilution and for arbitrary chemical compositions. It turns out
that, under certain conditions, two RP transitions are found (both
discontinuous). The ensuing intermediate phase is only isostatic right
on the limit with the floppy phase. Nevertheless, it can be argued
that this intermediate phase has a low density of redundancies and,
therefore, low self-stress.

This work is organized as follows. \Sec{sec:rp-cayley-trees} discusses
RP on Cayley Trees attached to a rigid boundary. Next in
\Sec{sec:rp-bethe-networks}, the analytical methods needed to solve RP
on homogeneous Bethe networks without a boundary are
reviewed. \Sec{sec:exact-calc-redund} discusses how the density of
redundant bonds is calculated analytically. \Sec{sec:analyt-expr-r}
contains an interpretation of the resulting analytic expressions,
showing that the system is isostatic at the transition point, while
\Sec{sec:use-r-locate} discusses how this expression is used to locate
the true transition point on a Bethe network, and considers some
specific examples for which analytic predictions are
derived. \Sec{sec:rigid-cons} generalizes the methods developed in
previous sections to the case of Mixed Bethe networks without chemical
order. A specific example is discussed, which displays two RP
transitions. Finally, \Sec{sec:discussion} contains a discussion of
results, some conclusions, and a perspective for future work.
\section{RP on Cayley Trees}
\label{sec:rp-cayley-trees}
\subsection{Self consistent equations}
Let us first review Central-Force RP on Cayley trees~\cite{MDLFRO97}
with coordination number $z$ (branching number $\alpha=z-1$), where
each bond is independently present with probability $b$.

Each node of the tree is assumed to be a rigid object with $g$ degrees
of freedom, i.e.~at least $g$ independent mechanical constraints
acting on it are needed in order to eliminate all its motions. In the
following, we will assume $g\geq 2$, as $g=1$ is usual
Percolation~\cite{MDCOR99}. Each present bond acts as a rotatable bar,
imposing one mechanical constraint to the degrees of freedom of the
two nodes it connects~\cite{CSR79}.

In a Cayley tree, rigidity stems from the boundary, which itself
constitutes layer zero ($l=0$) of the tree, and propagates away from
the boundary through present links.  Let $T_{l}$ be the probability
that an arbitrarily chosen node on layer $l$ be rigidly attached to
the boundary, when only its $\alpha$ downwards-pointing links are
considered. A node with this property is said to be
\emph{outwards-rigid} (o-rigid).

A node in layer $(l+1)$ will be o-rigid \emph{iif} it is connected to
$g$ or more o-rigid nodes in layer $l$.  Since each of its $\alpha$
links to neighbors on layer $l$ is independently present with
probability $b$, this node will be connected to $k$ o-rigid nodes in
layer $l$ with probability \hbox{$ P^{\alpha}_k(bT_{l}) \Def {\alpha
    \choose k} (bT_{l})^k (1-bT_{l})^{(\alpha-k)}$}. Therefore,
\begin{eqnarray}
\label{eq:1}
  T_{l+1}   = \sum_{k=g}^{\alpha} P^{\alpha}_k(bT_{l}).
\end{eqnarray}
Letting now $x_{l} = b T_{l}$ be the probability that a node in layer
$l$ be o-rigid \emph{and} connected to the node immediately above it
on layer $(l+1)$, one rewrites (\ref{eq:1}) as
\begin{equation}
  x_{l+1} =  b \sum_{k=g}^{\alpha} P^{\alpha}_k(x_{l}).
\end{equation}
Far away from the boundary (for $l \to \infty$), $x_l$ reaches a fixed
point $x$, that satisfies
\begin{equation}
  \label{eq:21}
  x =  b \sum_{k=g}^{\alpha} P^{\alpha}_k(x)  = b S_{g}^{\alpha}(x),
\end{equation}
where \hbox{$S_{l}^m(x)=\sum_{k=l}^m P_k^m(x)$}.  This self-consistent
equation defines the asymptotic rigid probability $T(b)$ as a function
of bond-density $b$, in the following way~\cite{MRPI03}: For $0 < x
\le 1$ calculate
\begin{equation} 
\left \{
\begin{array}{rcl}
 T(x) &=& S_{g}^{\alpha}(x)\\ 
 b(x) &=& x/T(x), \\ 
\end{array}
\right .
\label{eq:23}
\end{equation}
which together define $T(b)$ implicitly.  
\subsection{Multivaluedness of solutions}
Notice that $T(b)$, as given by (\ref{eq:23}), is multivalued (See red
lines in \Fig{fig:1}). The trivial solution $T=0 \forall b$ is always
stable. For $b>b_c^{Cayley}$ two additional solutions exist, one
stable and one unstable. Therefore, for $b_c^{Cayley} \leq b \leq 1$
there are two stable solutions $T(b)$ and therefore a physical
criterion is needed to choose the correct one. On a Cayley tree, a
well defined procedure exists to determine the appropriate solution,
which consists in iterating (\ref{eq:1}) from a given boundary value
$T_0$. If $T_0=1$, equations (\ref{eq:1}) give rise, for $g \geq 2$,
to a First-Order Critical RP Transition~\cite{MDLFRO97}, whereby the
order parameter $T$ displays a critical behavior superposed to a
first-order jump, above a critical bond density $b_c^{Cayley}$ (See
blue line in \Fig{fig:1} for an example).
\section{RP on Bethe Networks}
\label{sec:rp-bethe-networks}
A Bethe network has the same local structure as the Cayley tree
described in \Sec{sec:rp-cayley-trees}, but has no boundaries. Each
node of a Bethe network is connected to $z$ other nodes, without
exception. While a tree has no loops (except for those which run
through the boundary), any finite Bethe network is bound to have some
loops, but those are few in number and can be safely ignored.  Many
statistical models display the same behavior on Cayley trees with a
boundary as they do on Bethe networks without one. For RP this is not
the case.  It has been shown~\cite{DJTFMA99} that the 1st-order
critical behavior found for RP on Cayley trees~\cite{MDLFRO97} is an
artifact due to the excess constraints introduced by the rigid
boundary.  On a Bethe Network, equations (\ref{eq:23}) still
hold. However, in this case the RP transition is delayed and only
happens at a larger $b$ value, becoming a normal first-order
one~\cite{DJTFMA99}, as we review in the following section.
\subsection{Self-consistent  equations}
Consider a Bethe network with coordination number $z$ (branching
number $\alpha=z-1$), and $g$ degrees of freedom per site, where each
bond is present with probability $b$.  Choose an arbitrary site $i$ of
the network, and let $j$ be a randomly chosen neighbor of $i$. We will
say that $j$ is \emph{outwards-rigid} (o-rigid for short), if it is
rigidly attached to an infinite rigid cluster through
outwards-pointing links, i.e.~those not including the one that
connects $j$ to $i$. Precisely speaking, o-rigidity is not a property
of site $j$, but of site $j$ when considered from neighbor $i$. This
makes o-rigidity a property on the links of a directed graph. However,
this consideration is not essential here, and we will continue to
refer to o-rigidity as a property of neighboring \emph{sites}.

Now, let $T$ be the probability that a randomly chosen neighbor $j$ of
$i$ be o-rigid in the sense above. Notice that $j$ will be o-rigid
\emph{iif} it is connected to $g$ or more o-rigid neighbors, among its
$\alpha$ outwards neighbors. Node $j$ is connected to a neighbor that
is itself o-rigid, with probability $x=bT$. Therefore, equation
(\ref{eq:21}) holds for $x$ here as well. A little reflection allows
one to conclude that $T$ plays, on Bethe networks, a similar role as
on Cayley trees, except for the fact that, on Cayley trees, outwards
means ``towards the boundary''.

Notice that $T$ is \emph{not} the same as the density $R$ of rigid
nodes on the Bethe network. The density $R$ of rigid nodes is found as
follows.  A site will be rigidly connected to the infinite rigid
cluster \emph{iif} $g$ or more of its $z$ neighbors are o-rigid
\emph{and} connected to it.  Therefore
\begin{equation}
R = \sum_{j=g}^{z} P_j^{(z)}(x), 
\label{eq:14}
\end{equation}
where $x=bT$, and $T$, the probability for a neighbor to be o-rigid,
satisfies (\ref{eq:21}).

Once $T$ is known, $R$ derives from it trivially using (\ref{eq:14}),
so that its critical behavior derives from that of $T$. Therefore, we
will ignore $R$ and analyze $T$, in practice regarding it as the
``order parameter'' of the RP transition, although customarily it is
the density $R$ of the infinite cluster what plays that role.
\subsection{Multivaluedness of solutions}
The curve $T(b)$ derived from (\ref{eq:23}) is multivalued in general,
as already discussed for Cayley trees.  In the case of Bethe networks,
there is no boundary to determine the correct bulk solution by
propagation.  In order to identify the thermodynamically (globally)
stable solution on a Bethe network, one must introduce the concept of
redundant constraints and discuss their connection with a free-energy
for the RP problem. This has been done in detail previously and we
refer the reader to the relevant literature~\cite{DJTFMA99} for the
conceptual details.

For the sake of our work here, it suffices to say that the density of
redundant bonds $r(b)$ must satisfy two constraints: Firstly, it must
be everywhere continuous. In the second place, whenever multiple
(locally) stable solutions exist, the one with the largest $r$ is the
thermodynamically (globally) stable one. This second criterion
(maximization of $r$) depends on identifying the number of uncanceled
degrees of freedom in the system with a free-energy for the RP
problem. At present, this identification only has the status of a
reasonable hypothesis, i.e.~one that, having been tested in several
instances~\cite{DJTFMA99,MRPI03}, produces consistent
results. However, bear in mind that there is no formal proof of the
fact that the number of uncanceled degrees of freedom is the free
energy for RP. Establishing this formal connection constitutes one of
the most important results still awaiting completion in the field of
RP.

It turns out that the density of redundant bonds $r(p)$ results from
integrating the density of overconstrained bonds, which, in
mean-field, is simply $T^2$.  Therefore,
\begin{equation}
  r(b) = \int_0^p T^2(b) db.
\label{eq:26}
\end{equation}
This expression was used recently~\cite{DJTFMA99} to find $r(b)$ by
\emph{numerical integration} of $T^2(b)$, while $T(b)$ was calculated
by starting from $x=1$ and iterating (\ref{eq:21}), for a given
link-density $b$.

However, as we now show, $r(b)$ can be calculated analytically with
little effort. This will allow us to locate the Bethe RP transition
exactly. A similar calculation was done previously~\cite{MRPI03}, but
only for the (somewhat simpler) case of \ER\ graphs. The following
section describes how $r(b)$ can be calculated analytically for Bethe
networks, starting from (\ref{eq:26}), and illustrates its use in
exactly locating the RP transition with a few simple examples.
%%%%%%%%%%%%%%%%%%%%%%%%%%%%%%%%%%%%%%%%%%%%%%%%%%%%%%%%
\begin{figure}[!h]  
\centerline{\textbf{\Large a)} 
\psfig{figure=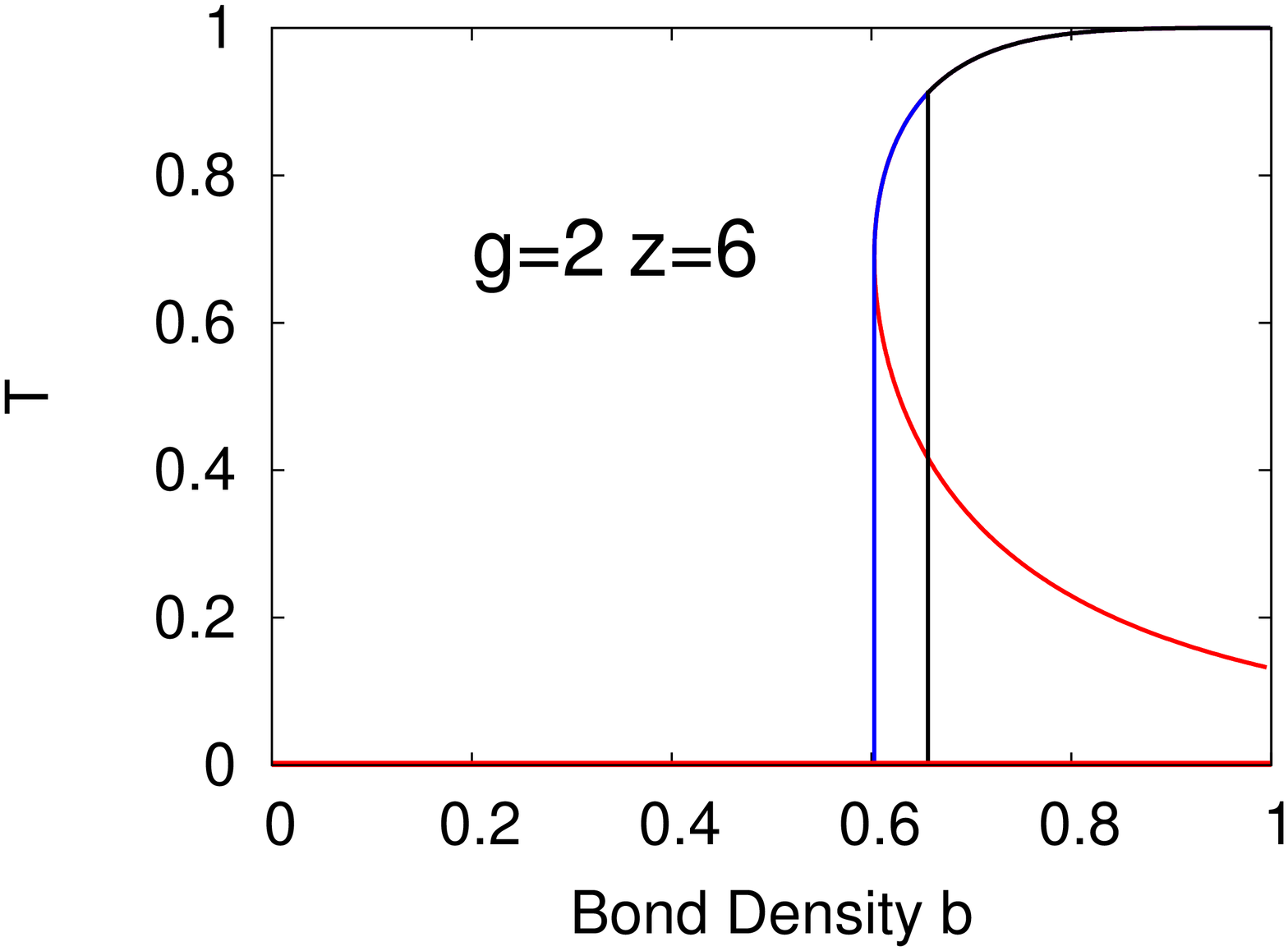,width=0.95\linewidth,angle=270}
}\centerline{ \textbf{\Large b)} 
\psfig{figure=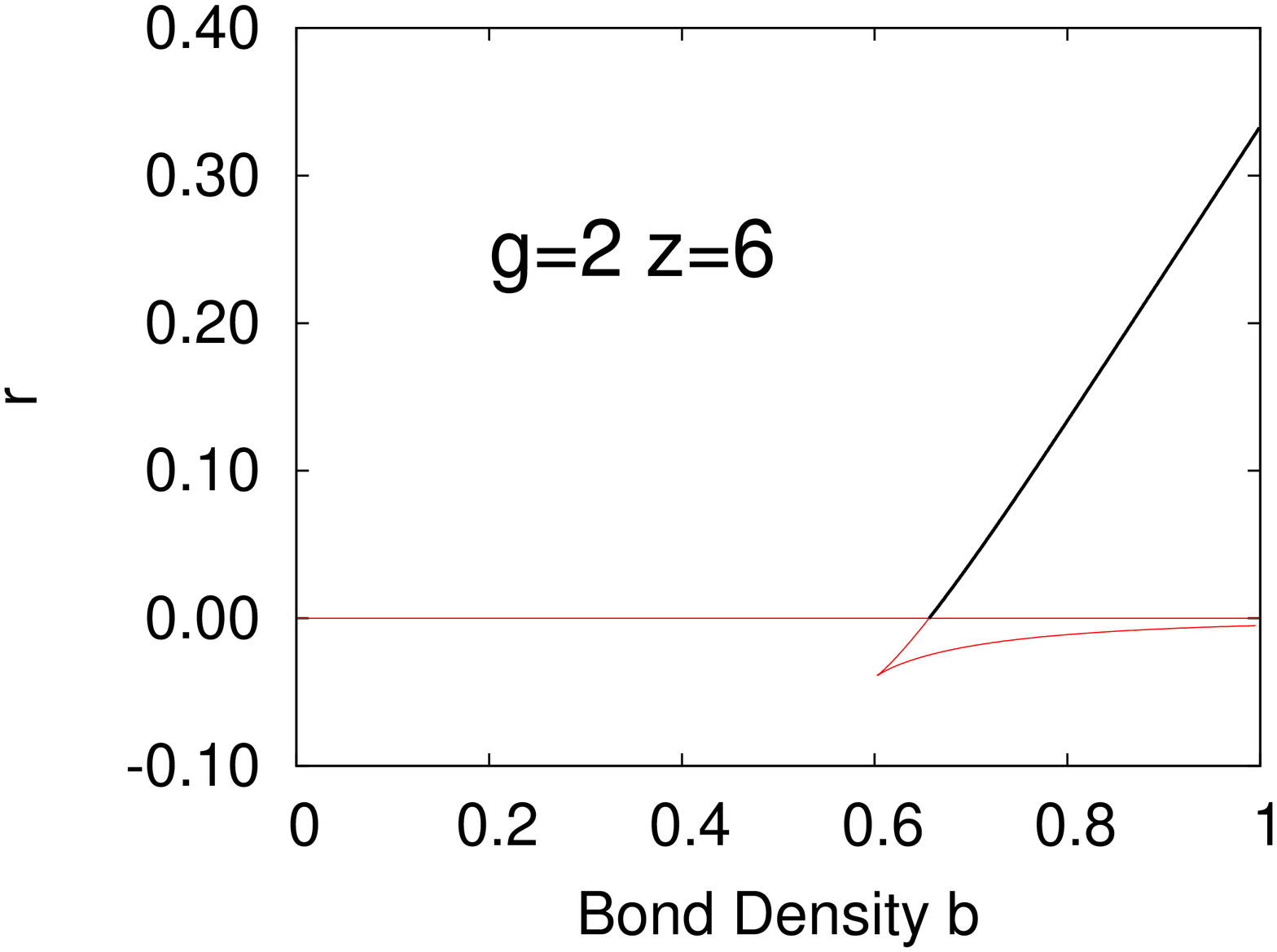,width=0.95\linewidth,angle=270}
}
\caption{\textbf{a)} Red lines (partially hidden below black and blue
  ones) indicate the multivalued solution $T(b)$ obtained from
  equations (\ref{eq:23}) with $0\leq x\leq 1$, for a Cayley tree with
  $g=2$ and $z=6$. The entire branch \hbox{$T=0 \forall b$} actually
  corresponds to a single point $x=0$, and is always a locally stable
  solution. This branch is continued, for $x>0$, by an unstable
  branch with $b$ a decreasing function of $x$, up to the reversal
  point $b_c^{Cayley}$, where it becomes locally stable, with $b$ an
  increasing function of $x$.  On a Cayley tree (blue line), a
  First-Order Critical RP transition happens at bond density
  \hbox{$b_c^{Cayley}=0.602789$} if the boundary is perfectly
  rigid. On a Bethe network with the same $g$ and $z$ the transition
  happens later (black line), becoming an ordinary First-Order RP
  transition. The exact location \hbox{$b_c^{Bethe}=0.656511134$} of
  the transition on the Bethe network can be determined analytically
  (See (\ref{eq:bethe0})) from the requirement that the density $r$ of
  redundant bonds (\textbf{b)}) be continuous~\cite{DJTFMA99}.}
\label{fig:1}
\end{figure}
%%%%%%%%%%%%%%%%%%%%%%%%%%%%%%%%%%%%%%%%%%%%%%%%%%%%%%%%
\subsection{Exact calculation of redundant bond density}
\label{sec:exact-calc-redund}
\subsubsection{Analytic expression for $r$}
\label{sec:analyt-expr-r}
Given that $x=bT$, one has that $T^2 db = Tdx-xdT$, and therefore
\begin{equation}
  r(b) = \int_0^{x(b)} T dx - \int_0^{T(b)} x dT.
\end{equation}
The second integral on the r.h.s.~is done by parts to obtain
\begin{equation}
  r(b) = 2\int_0^x T(x) dx - \left . x T \right |_0^x.
\label{eq:18}
\end{equation}
Using (\ref{eq:23}), the remaining integral can be calculated
exactly:
\begin{eqnarray} \nonumber
\int dx  T(x) &=& \sum_{j=g}^{\alpha} \int P_j^{\alpha}(x) dx =
\frac{1}{z} \sum_{k=g}^{\alpha} \sum_{j=k+1}^{z} P^{z}_j(x) \\
&=& \frac{1}{z}  \sum_{j=g+1}^{z} (j-g) P^{z}_j(x).
\end{eqnarray}
Therefore,
\begin{equation}
r(x)= \frac{2}{z}  \sum_{j=g+1}^{z} (j-g) P^{z}_j(x) 
-  \sum_{j=g}^{\alpha} x P_j^{\alpha}(x).
\end{equation}
Taking into account that \hbox{$(j+1) P^{z}_{j+1}(x) = z x
  P_j^{\alpha}(x)$}, one can also write
\begin{equation}
  r(x)= \frac{1}{z} \sum_{j=g+1}^z (j-2g) P_j^{z}(x), 
%sort of trivial% =   \frac{2}{z} \sum_{j=g+1}^z (\frac{j}{2}-g) P_j^{z}(x),
\label{eq:27}
\end{equation}
which also defines $r(b)$ implicitly through (\ref{eq:23}).  

Expression (\ref{eq:27}) can be interpreted as follows: Among rigid
sites, i.e.~those which have $g$ or more rigid neighbors, those with
exactly $g$ rigid neighbors contribute no redundant bonds. This is so
because sites with exactly $g$ rigid neighbors are
\emph{isostatically} connected to the infinite rigid cluster, and can
be removed without altering the rigid properties of the rest of the
system.  These sites are similar to \emph{dangling ends} in scalar
percolation~\cite{SAITP94}.

Noticing that $r$ is a link-density, by examination of (\ref{eq:27})
one also concludes that sites with $j>g$ o-rigid neighbors contribute
with \hbox{$(j/2-g)$} redundancies per site, which is a consequence of
the fact that each link is shared by two nodes.  In the following,
sites with more than $g$ o-rigid neighbors will be called
\emph{backbone rigid sites}. They form the $(g+1)$-core contained
within the rigid cluster~\cite{MRPI03}.

Notice that \hbox{$(j/2-g)$} in (\ref{eq:27}) is negative unless
$z\geq 2g$.  Since $r$ has to be nonnegative, we conclude that $z \geq
2g$ is a necessary condition for rigidity on a Bethe network.  This is
consistent with a simple dof counting argument.  On the other hand,
rigidity on a Cayley tree only requires $z \geq g+1$, because the
boundary provides extra constraints.
\subsubsection{Use of $r$ to locate the transition}
\label{sec:use-r-locate}
The continuity requirement for $r$ determines the true transition
point on a Bethe network~\cite{DJTFMA99} as follows: Below the
critical point, the order parameter $T$, and thus $x$, are identically
zero.  The density of redundant bonds is then zero at the critical
point, so 
\begin{equation}
r(x_c) = \frac{1}{z} \sum_{j=g+1}^z (j-2g) P_j^{z}(x_c) = 0.
\label{eq:bethe0}
\end{equation}
Equation (\ref{eq:bethe0}) determines $x_c$ and, therefore, $b_c$.
This can also be written as
\begin{equation} 
\frac{\sum_{j=g+1}^z  j P_j^{z}(x_c) }
 {\sum_{j=g+1}^z  P_j^{z}(x_c)} = 2g,
\label{eq:2}
\end{equation}
showing that the average number of o-rigid neighbors of a backbone
site is exactly $2g$ at the rigidity transition.  This means that,
when only backbone sites in the infinite rigid cluster are considered,
constraint balance is satisfied at the transition point. Sites which
are isostatically connected to the rigid cluster (those with exactly
$g$ o-rigid neighbors) also have exact constraint balance because
their links are not shared between two o-rigid sites (since they are
themselves not o-rigid).  We thus conclude that the entire rigid
cluster (backbone sites plus isostatically connected sites) satisfies
constraint balance at the transition point. This is, of course, just a
consequence of the fact that the density of redundancies is zero at
the transition point ($r(x_c)=0$).

However, bear in mind that our calculation neglects finite-size
effects, so that $r=0$ only means that the number of redundancies is
sub-extensive. Clearly, the number of redundancies in the system
cannot be exactly zero at the transition point, since a large fraction
$T^2$ of the links are overconstrained
there~\footnote{However, this does not necessarily imply
  that the number of redundancies is large. A single redundant link
  may be enough to make a large number of links
  overconstrained.}. Now, the precise definition of
\emph{isostaticity} requires that a system be rigid and have no
redundant constraints at all. By a slight abuse of language, we can
extend this definition to include systems that are rigid with a
sub-extensive number of redundancies.  Accepting this extended
definition, we can then say that the rigid cluster is \emph{isostatic}
at the transition point in this mean-field system.

The general procedure to calculate $b_c$ consists in first finding
$x_c$ from (\ref{eq:bethe0}), then calculating $b_c = x_c/T(x_c)$ with
$T(x)$ given by (\ref{eq:23}).  For small values of $g$ and $z$, the
roots of (\ref{eq:bethe0}) can be found analytically.  For $g=2$ and
$z=4$, the sum in (\ref{eq:bethe0}) only contains one term giving rise
to $x_c=1$. This implies $T(x_c)=1$ and therefore $b_c=1$. Thus, a
Bethe network with $g=2$ and $z=4$ is only rigid if undiluted, as
discussed already. On the other hand, the Cayley tree with $g=2$ and
$z=4$ \emph{can} be rigid even in the presence of some small amount of
dilution (i.e.~$b_c^{Cayley}<1$).  For $g=2$ and $z=5$, the
corresponding quadratic equation results in $b_c=0.83484234$.  For
$g=2$ and $z=6$ a cubic equation is obtained, and its solution results
in $b_c=0.656511134$.  This last value is consistent with, but more
precise than, $b_c=0.656$, as obtained numerically in previous
work~\cite{DJTFMA99} with the help of matching algorithms for
RP~\cite{MAEA96,JHAAF97}. Notice that $b_c$ is slightly smaller than
$2/3$, which would result for GCB for this case.

\Fig{fig:1} shows this last example with $g=2$ and $z=6$, both for
Cayley trees (blue line) and for Bethe networks (black line). The
results for Bethe in this figure where obtained by calculating $T(x),
b(x)$ and $r(x)$ as given by \Eqns{eq:23} and (\ref{eq:27}) (red
lines), and then choosing as correct solution (black lines) the branch
that provides the largest value for $r(x)$.

Notice that the density of redundant bonds makes a loop-shaped graph
limited by two return points (\Fig{fig:1}b, red lines), which
correspond to spinodals in the $T(b)$ curve. Actually, one of the two
spinodals in the example of \Fig{fig:1} is located at $\infty$, but
the addition of a small field~\cite{MRPI03} can bring it to finite
values of $b$. The process by which this loop is removed by choosing
the branch that maximizes $r$ is a \emph{Maxwell
  construction}~\cite{DJTFMA99}.
\section{RP on mixed Bethe networks}
\label{sec:rigid-cons}
Although more complex networks can be readily analyzed with the
methods developed in this work, the Mixed Bethe Networks (MBN) studied
here only have two types $\nu=1,2$ of sites. A fraction $p_1$ of the
sites has coordination $z_1$ and $g_1$ dof, while the remaining
fraction $p_2=1-p_1$ has coordination $z_2$ and $g_2$ dof. We will
refer to $p_1$ as the ``chemical composition'' of the network.

Consider an undiluted such network with given values of $p_1, z_1$ and
$z_2$.  The probability $p_{\nu \rho}$ that a randomly chosen neighbor
of a site of type $\nu$ be of type $\rho$ depends in general upon the
degree to which links between like atoms are favored or disfavored.
Under the hypothesis of random bonding, i.e.~when all links are
equally probable, one has that $p_{\nu \rho}$ only depends on $\rho$:
\begin{equation}
  p_{\nu \rho} = \frac{z_\rho p_\rho}{z_1 p_1+z_2p_2} \Def   \peff_\rho, \quad
\hbox{for} \quad
\nu,\rho=1,2.
\label{eq:29}
\end{equation}
The general case of RP on MBNs with chemical order, i.e.~when certain
types of links are favored, can also be treated with the methods
described in this work, but the mathematics becomes more
complicated~\footnote{C.~Moukarzel to be published.}, and will not be
discussed in this first presentation.

In this work we consider randomly bonded MBNs of the type described
above, whose links have been randomly removed, or diluted. For
simplicity, we assume homogeneous dilution, i.e.~all bonds are present
with probability $b$, no matter what types of sites they link.
\subsection{Self-consistent equations for rigid probability}
\label{sec:self-cons-equat}
Let $T_\rho,$ with $\rho=1,2$, be the probability that a neighbor,
provided it is of type $\rho$, be outwards-rigid.  An arbitrarily
chosen neighbor of a central site, no matter its type, is then
outwards- rigid \emph{and} connected to the central site, with
probability \hbox{$x=b(\peff_1 T_1 + \peff_2 T_2)$}.

A neighbor of type $\rho$ must in turn have $g_\rho$ or more o-rigid
and connected outwards-neighbors (of any type) in order to be itself
outwards-rigid with respect to the central site.  Therefore, the
outwards-rigid probabilities $T_\rho$ of neighbors satisfy
\begin{equation}
  T_\rho = \sum_{k=g_\rho}^{\alpha_\rho} {\alpha_\rho\choose k} x^k (1-x)^{\alpha_\rho-k} = 
  S_{g_\rho}^{\alpha_\rho}(x),
\label{eq:12}
\end{equation}
where $S_{g}^{\alpha}(x)$ is the same as defined in (\ref{eq:21}), and
\hbox{$\alpha_\rho=z_\rho-1$}.  

The self-consistent equations defining rigidity on an MBN with no
chemical order (random bonding) are then
\begin{eqnarray}
\label{eq:33}
T_1 &=&    S_{g_1}^{\alpha_1}(x) \\
\label{eq:34}
T_2 &=&    S_{g_2}^{\alpha_2}(x) \\
\label{eq:15}
x &=& b (\peff_1 T_1 + \peff_2  T_2) \Def b T.
\end{eqnarray}
Notice that rigidity on randomly-bonded MBNs depends on two
parameters: the link density $b$ and the chemical composition
$p_1$. The system's phase diagram then results from analyzing the
average outwards-rigid probability $T(b,p_1)$ as a function of these
two parameters.  We will next discuss, in this phase space,
trajectories at fixed $b$ (the driving parameter is $p_1)$ or at fixed
$p_1$ (the driving parameter is $b$).  In either case, $T$ can be
calculated exactly, as described in the following.
% %%%%%%%%%%%%%%%%%%%%%%%%%%%%%%%%%%%%%%%%%%%%%%%%%%%%%%%%
\begin{figure}[!h]  
\centerline{ \textbf{\Large a)} 
\psfig{figure=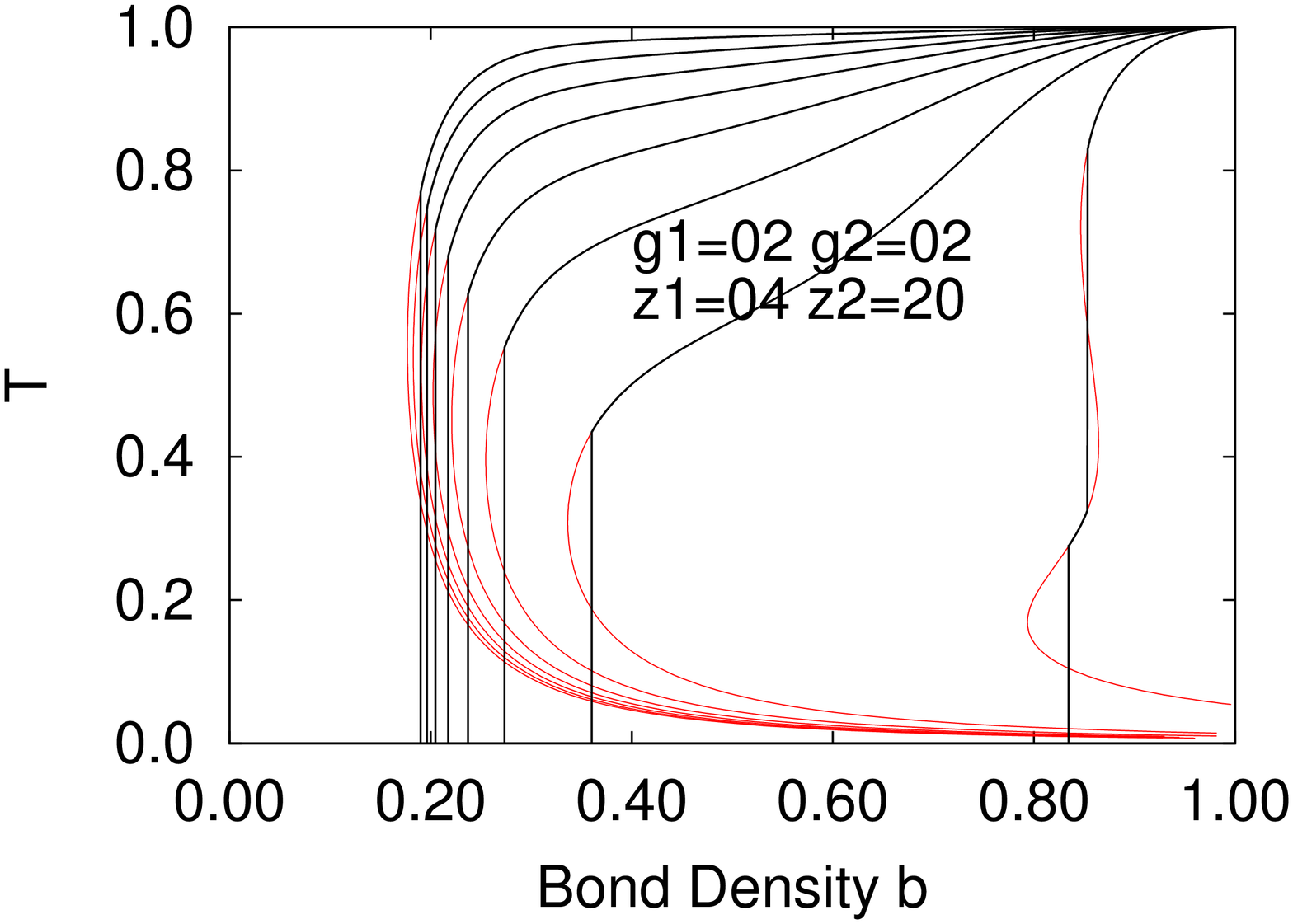,width=0.82\linewidth,angle=270}
} 
\centerline{ \textbf{\Large b)} 
\psfig{figure=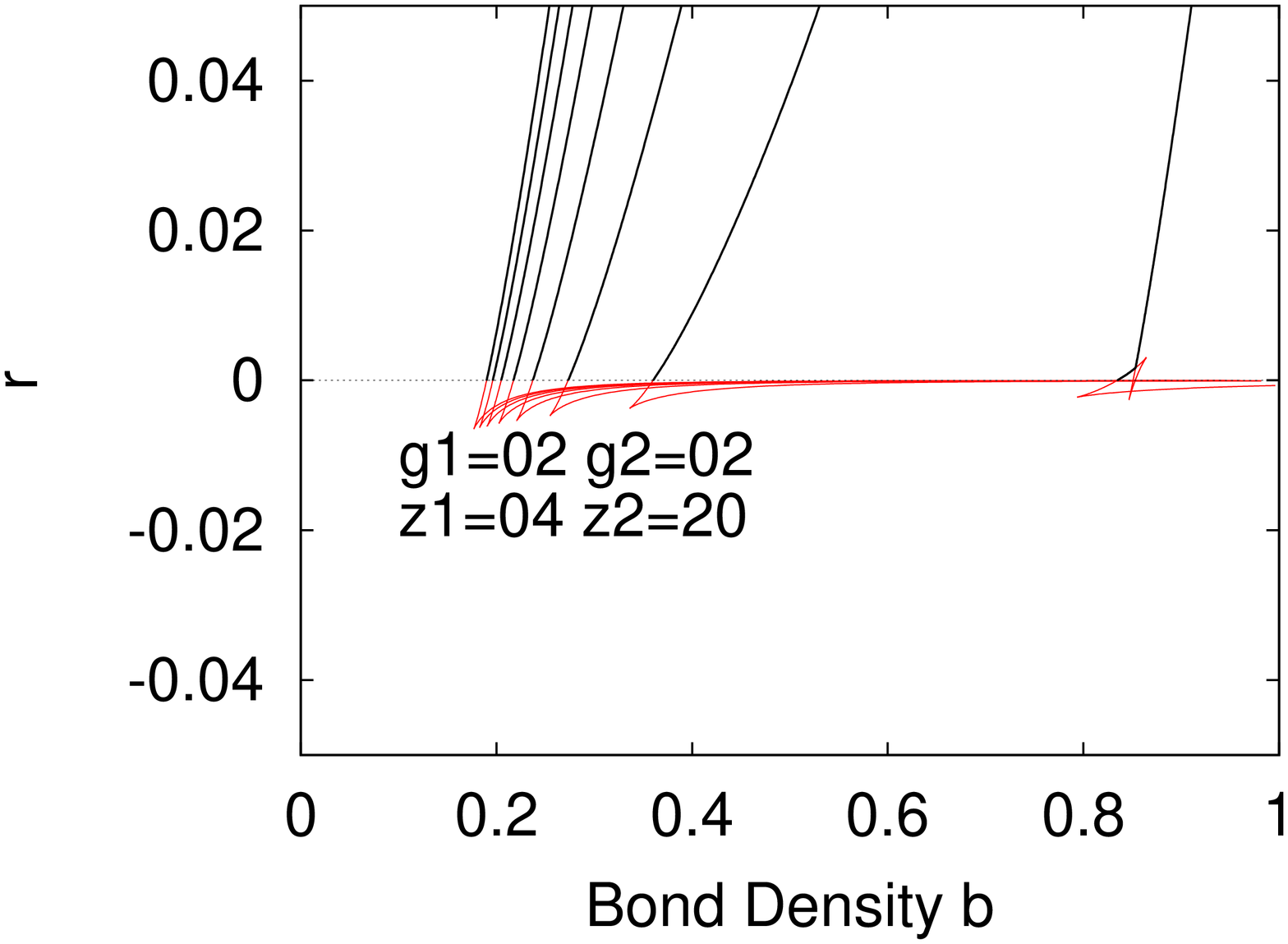,width=0.92\linewidth,angle=270}
} 
\caption{Example of $b$-driven phase transitions on mixed Bethe
  networks with $g_1=g_2=2$, $z_1=4$, and $z_2=20$. The ``weak''
  fraction $p_1$ is, from left to right:
  \hbox{$p_1=0.12,0.24,0.36,\ldots,0.96$}.  \textbf{a)} Average
  o-rigid probability $T$, defined by (\ref{eq:15}), as a function of
  link-density $b$. The leftmost case shows two phase transitions.
  \textbf{b)} Density or redundant links $r$, as given by
  (\ref{eq:rfinal}), as a function of link-density $b$. Notice the
  double Maxwell-loop in the rightmost case.  }
\label{fig:vsgamma}
\end{figure}
% %%%%%%%%%%%%%%%%%%%%%%%%%%%%%%%%%%%%%%%%%%%%%%%%%%%%%%%%
% %%%%%%%%%%%%%%%%%%%%%%%%%%%%%%%%%%%%%%%%%%%%%%%%%%%%%%%%
\begin{figure}[!h]  
\centerline{ \textbf{\Large a)} 
\psfig{figure=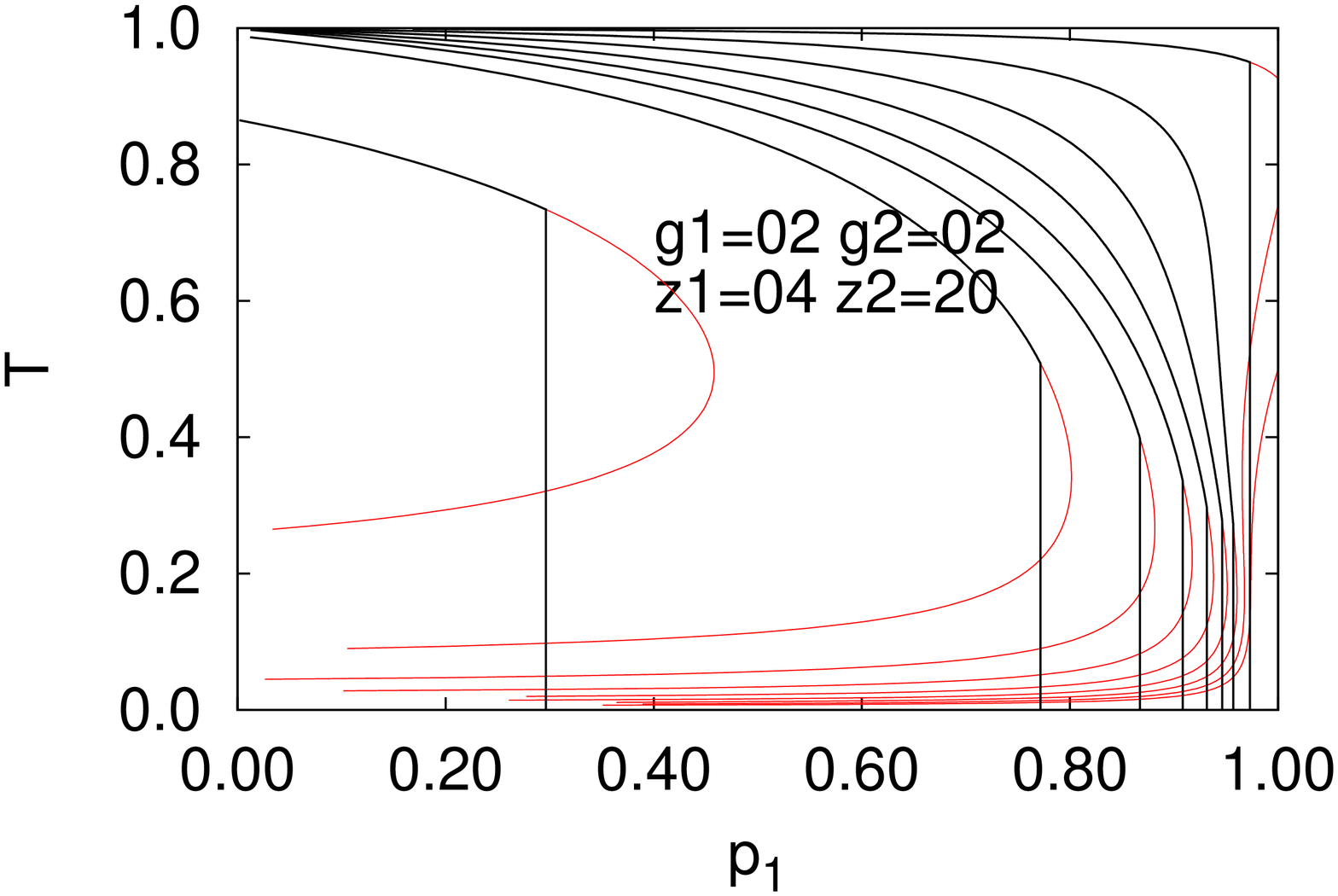,width=0.82\linewidth,angle=270}
} 
\centerline{ \textbf{\Large b)} 
\psfig{figure=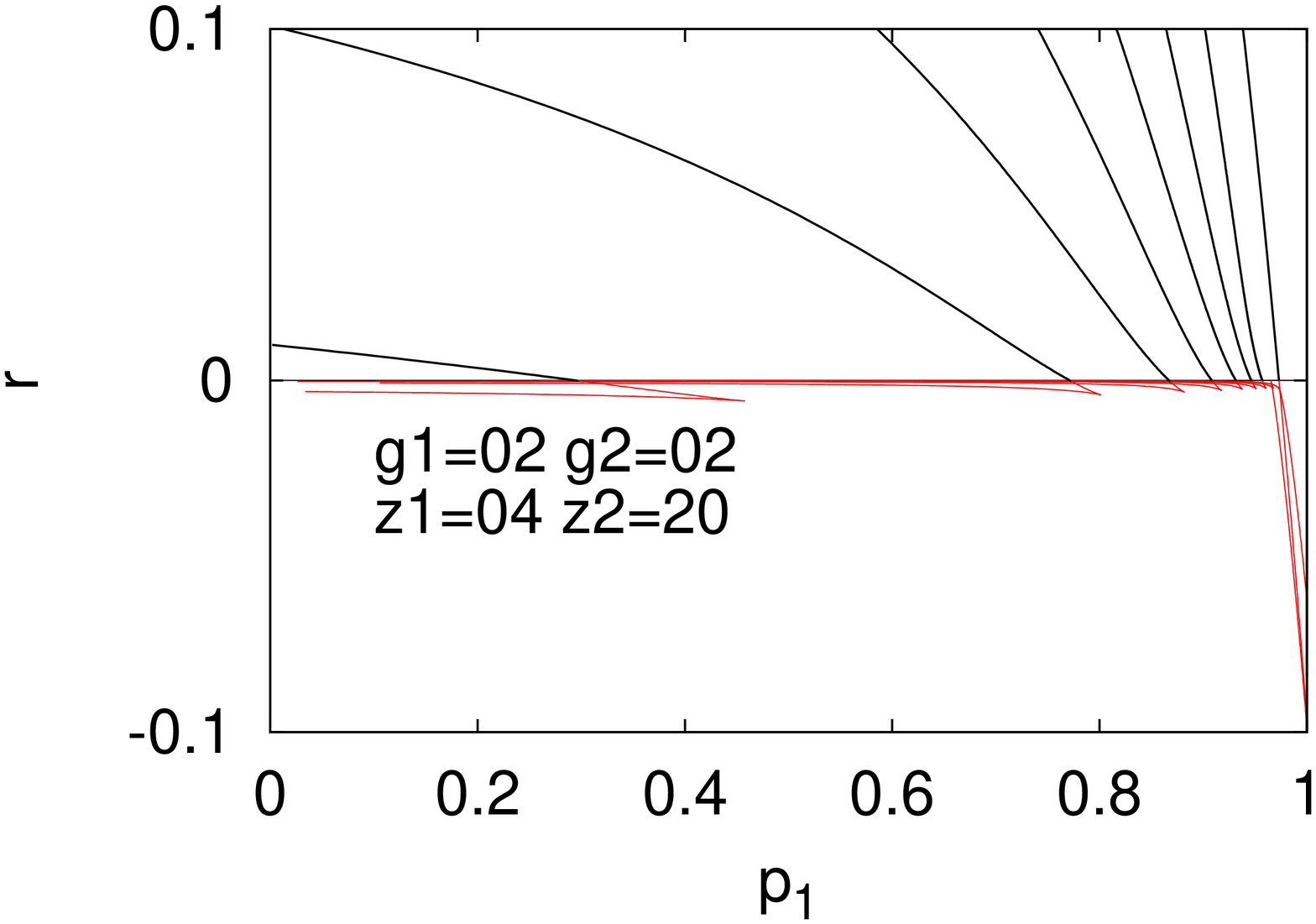,width=0.92\linewidth,angle=270}
} 
\caption{Example of $p_1$-driven phase transitions on mixed Bethe
  networks with $g_1=g_2=2$, $z_1=4$, and $z_2=20$. The link-density
  is, from left to right: \hbox{$b=0.2, 0.3,0.4,\ldots$} \textbf{a)}
  Average o-rigid probability $T$, defined by (\ref{eq:15}), as a
  function of chemical composition $p_1$. Double phase transitions,
  which happen for $b\approx 0.85$ are not seen in this figure. They
  are displayed in detail in \Fig{fig:2vsp1}.  \textbf{b)} Density or
  redundant links $r$, as given by (\ref{eq:rfinal}), as a function of
  chemical composition $p_1$.  }
\label{fig:vsp1}
\end{figure}
% %%%%%%%%%%%%%%%%%%%%%%%%%%%%%%%%%%%%%%%%%%%%%%%%%%%%%%%%
\subsubsection{Connectivity $b$ drives the transition}
Assume that $p_1$ and $p_2=1-p_1$ are given, so we work on MBNs with
fixed chemical composition and varying connectivity $b$. First, from
$p_1$ and $p_2$ find $\peff_1$ and $\peff_2$ using (\ref{eq:29}).
Then, for \hbox{$0\leq x \leq 1$}, calculate
\begin{equation}
\left \{
\begin{array}{rcl}
T_1 &=& S_{g_1}^{\alpha_1}(x) \\ 
T_2 &=& S_{g_2}^{\alpha_2}(x), \\
T &=&  ( \peff_1 T_1 + \peff_2 T_2 ),
\end{array}
\right .
\end{equation}
and then find $b(x)$ as
\begin{equation}
b =  \frac{x}{( \peff_1 T_1 + \peff_2 T_2 )} = \frac{x}{T} .
\label{eq:16}
\end{equation}
This procedure gives all branches of the eventually multivalued curve
$T(b)$ for a given prespecified chemical composition $p_1$. See red
lines in \Fig{fig:vsgamma} for an example. The true solution (black
lines in \Fig{fig:vsgamma}) is determined by the continuity of the
density of redundant bonds $r$, as discussed in
\Sec{sec:continuity-r}.
\subsubsection{Chemical composition $p_1$ drives the transition}
In this case we assume $b$ to be known, i.e.~work on graphs with
\emph{fixed dilution} and varying chemical composition $p_1$.  For
$0\leq x \leq 1$ we first calculate:
\begin{equation} 
\left \{
\begin{array}{rcl}
T_1 &=& S_{g_1}^{\alpha_1}(x) \\ 
T_2 &=& S_{g_2}^{\alpha_2}(x),
\end{array}
\right .
\end{equation}
and then use (\ref{eq:15}) in the form
\begin{equation}
\peff_1 = \frac{x/b - T_2}{T_1-T_2}, 
\end{equation}
to find $\peff_1(x)$. From $\peff_1$ one can find the chemical
composition, which is given by 
\begin{equation}
p_1 = \peff_1 z_2/(z_2 \peff_1 +  z_1 (1-\peff_1))
\label{eq:17}
\end{equation}
This procedure gives all branches of the
eventually multivalued curve $T(p_1)$, i.e.~rigidity as a function of
chemical composition, for a given prespecified bond dilution $b$. See
red lines in \Fig{fig:vsp1} for an example. The true solution (black
lines in \Fig{fig:vsp1}) is determined by the continuity of the
density of redundant bonds $r$, as discussed in
\Sec{sec:continuity-r}.
% %%%%%%%%%%%%%%%%%%%%%%%%%%%%%%%%%%%%%%%%%%%%%%%%%%%%%%%%
\begin{figure}[!h]  
\centerline{ \textbf{\Large a)} 
\psfig{figure=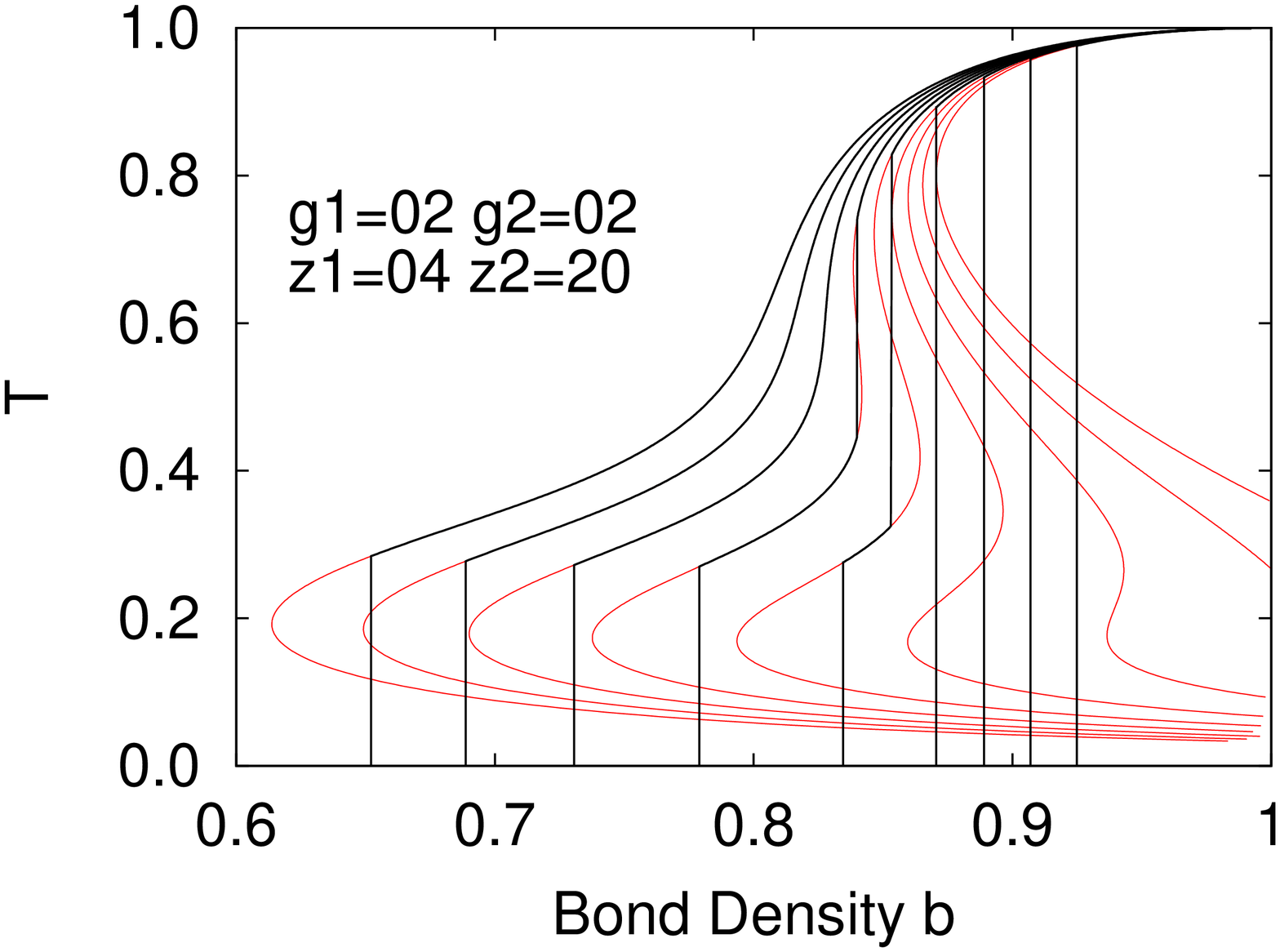,width=0.92\linewidth,angle=270}
} 
\centerline{ \textbf{\Large b)} 
\psfig{figure=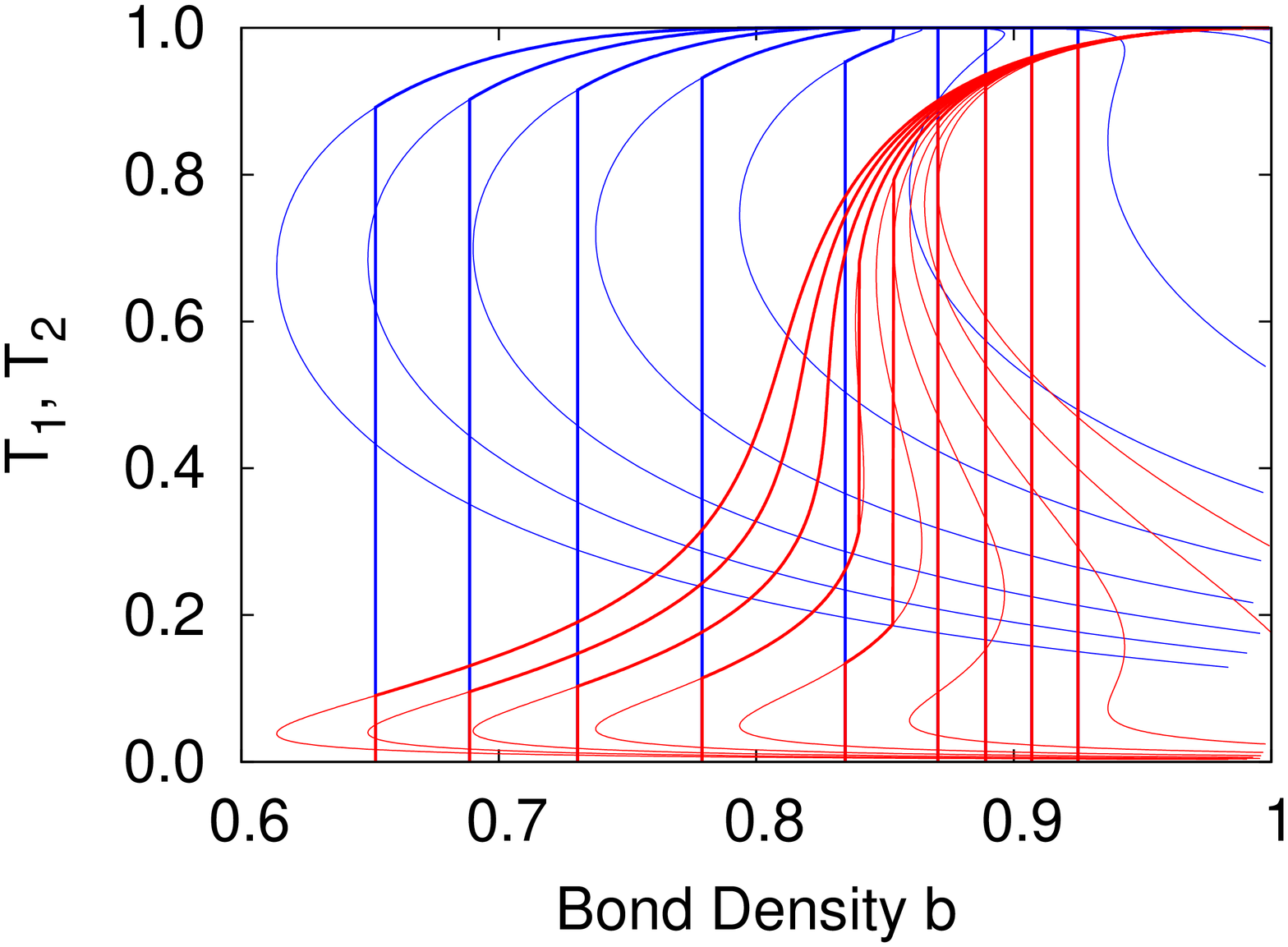,width=0.92\linewidth,angle=270}
} 
\centerline{ \textbf{\Large c)} 
\psfig{figure=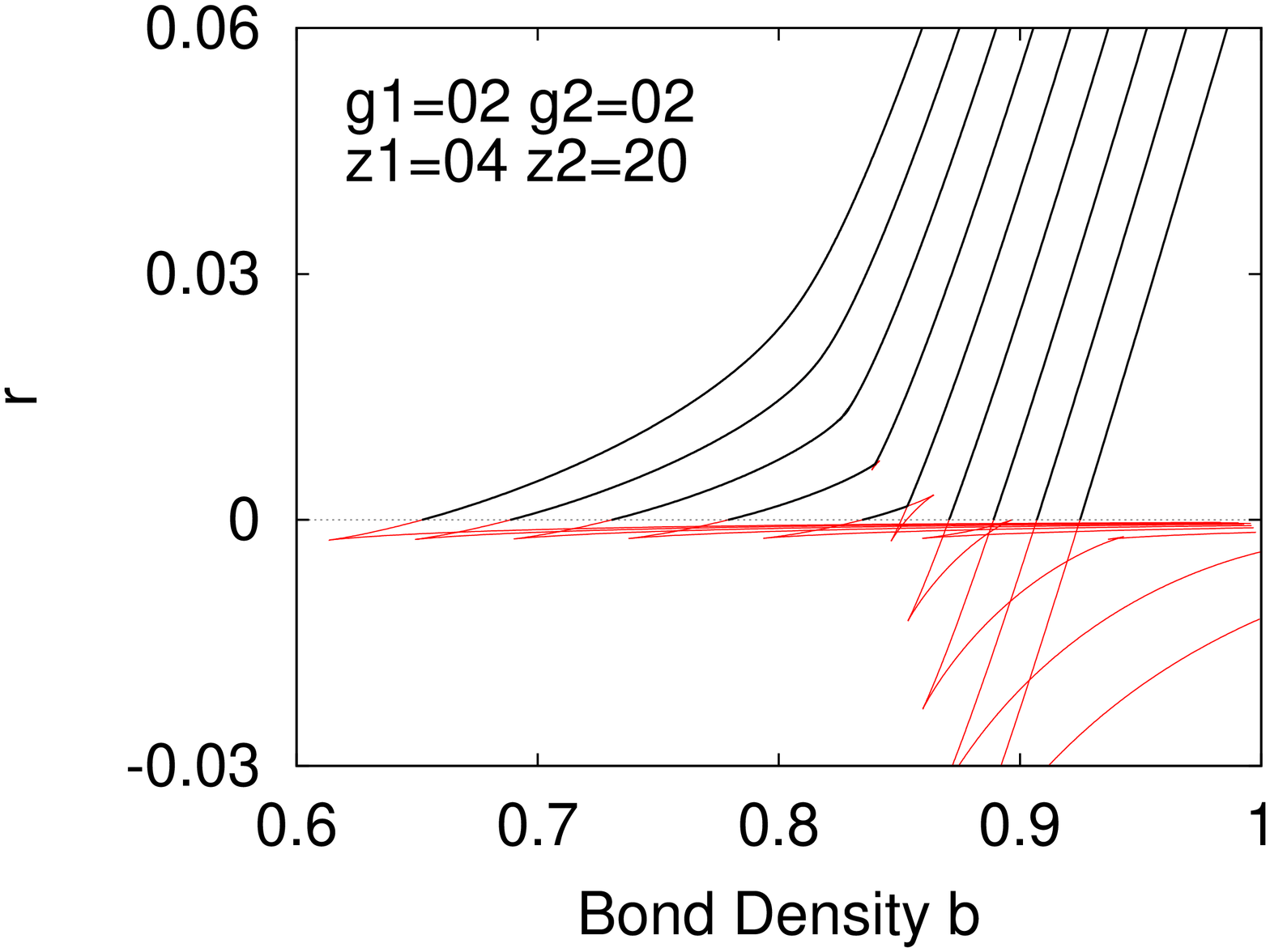,width=0.92\linewidth,angle=270}
} 
\caption{Zoom of double phase transitions in the $b$-driven case, on
  mixed Bethe networks with $g_1=g_2=2$, $z_1=4$, and $z_2=20$. The
  ``weak'' fraction $p_1$ is, from left to right:
  \hbox{$p1=0.940,0.945,0.950,\ldots,0.980$} \textbf{a)} Average
  o-rigid probability $T$, defined by (\ref{eq:15}).  \textbf{b)}
  o-rigid probabilities for each of the components.  \textbf{c)}
  Density or redundant links $r$, as given by (\ref{eq:rfinal}).}
\label{fig:2vsgamma}
\end{figure}
% %%%%%%%%%%%%%%%%%%%%%%%%%%%%%%%%%%%%%%%%%%%%%%%%%%%%%%%%
% %%%%%%%%%%%%%%%%%%%%%%%%%%%%%%%%%%%%%%%%%%%%%%%%%%%%%%%%
\begin{figure}[!h]  
\centerline{ \textbf{\Large a)} 
\psfig{figure=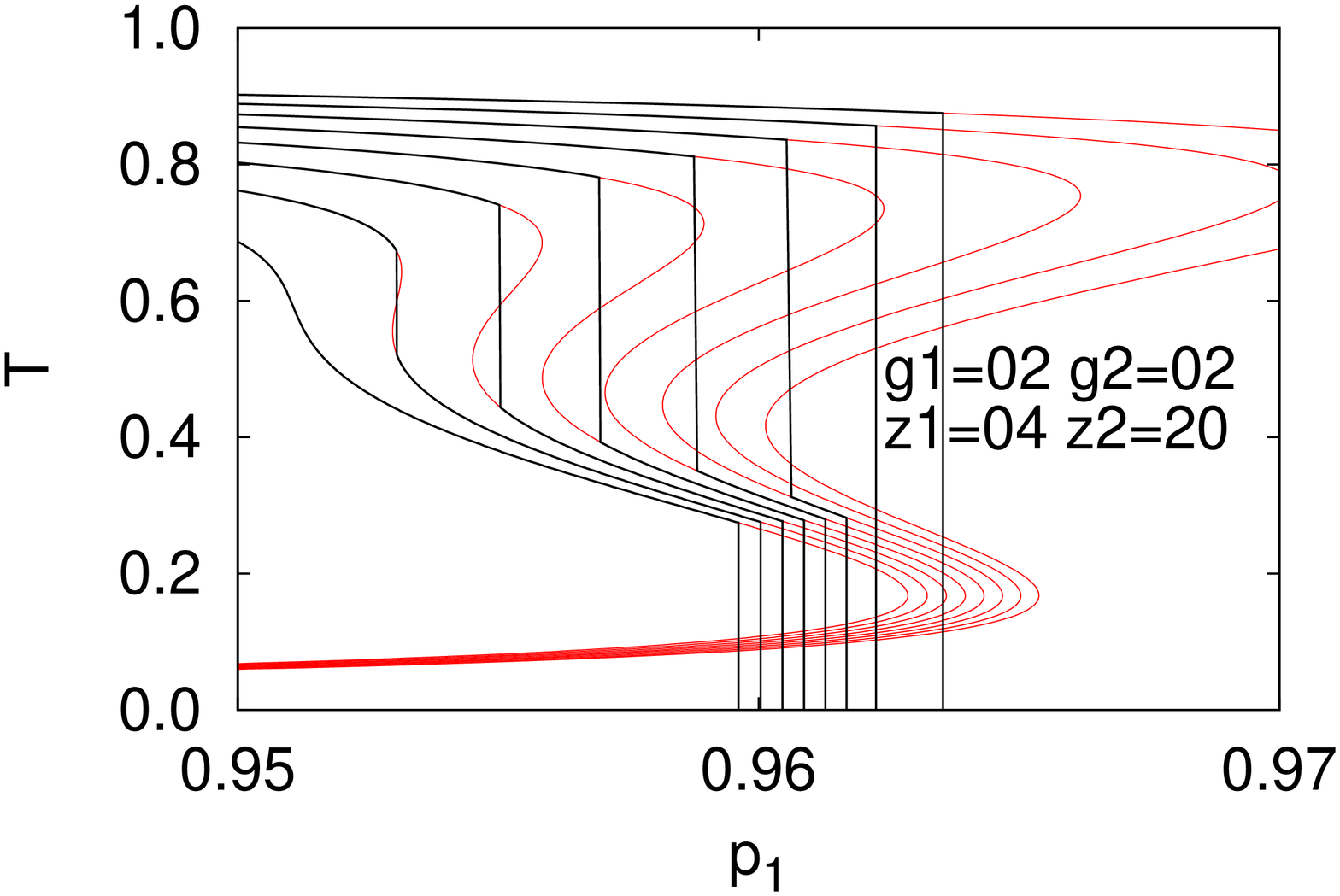,width=0.92\linewidth,angle=270}
} 
\centerline{ \textbf{\Large b)} 
\psfig{figure=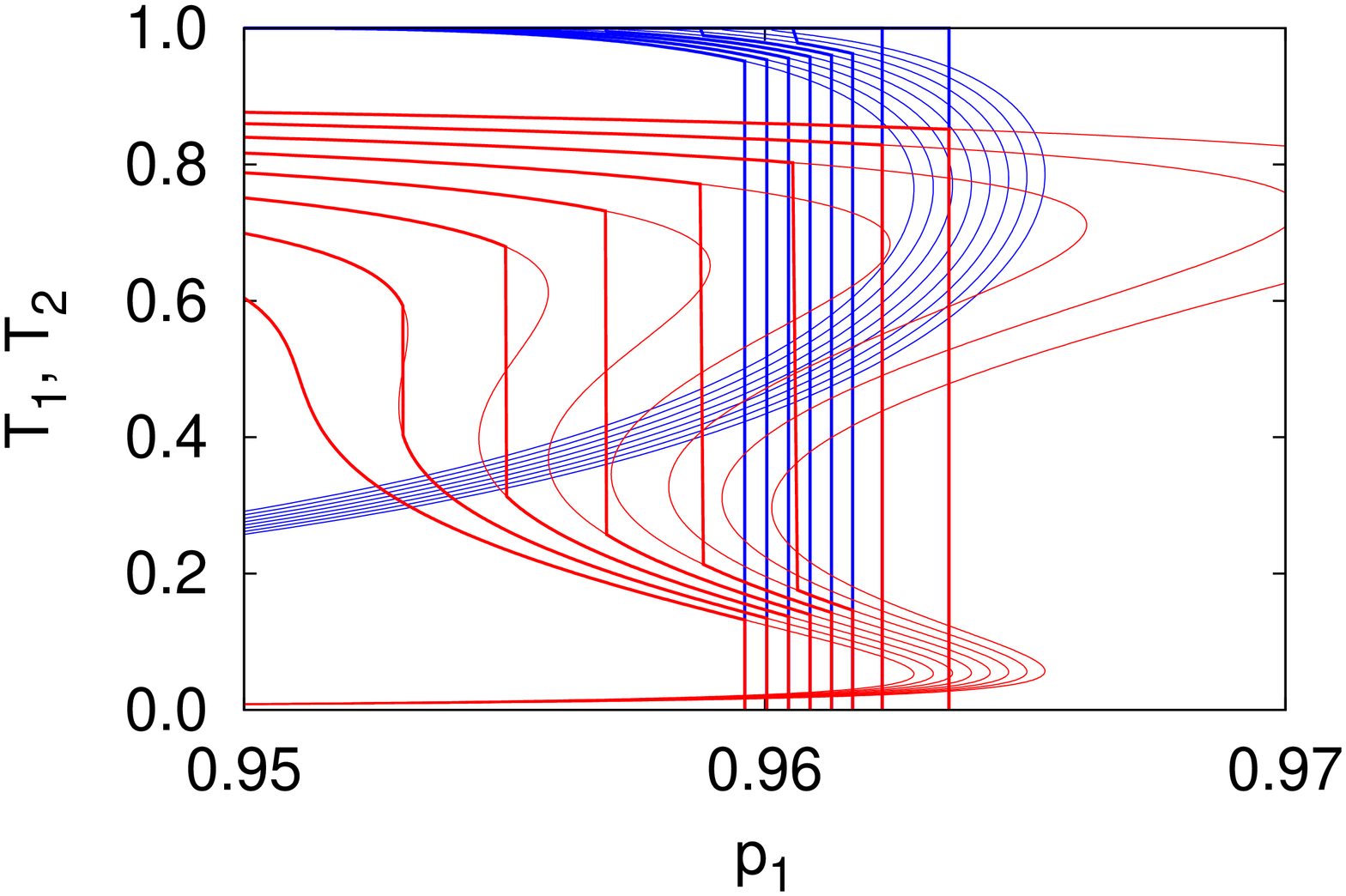,width=0.92\linewidth,angle=270}
} 
\centerline{ \textbf{\Large c)} 
\psfig{figure=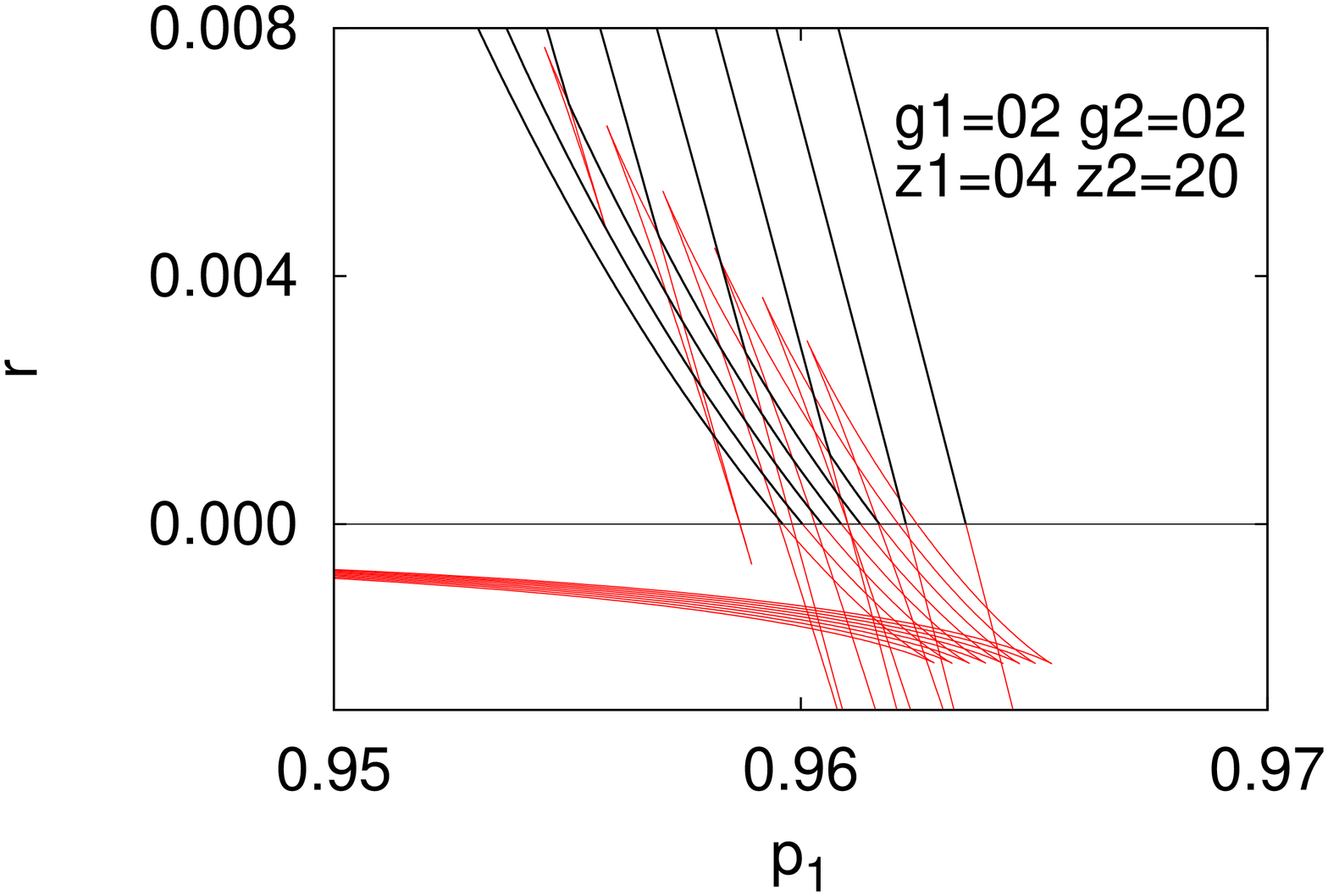,width=0.92\linewidth,angle=270}
} 
\caption{Zoom of double phase transitions in the $p_1$-driven case, on
  mixed Bethe networks with $g_1=g_2=2$, $z_1=4$, and $z_2=20$. The
  link-density $b$ is, from left to right:
  \hbox{$b=0.830,0.835,0.840,\ldots,0.865$}.  \textbf{a)} Average
  o-rigid probability $T$, defined by (\ref{eq:15}).  \textbf{b)}
  o-rigid probabilities for each of the components.  \textbf{c)}
  Density or redundant links $r$, as given by (\ref{eq:rfinal}).}
\label{fig:2vsp1}
\end{figure}
% %%%%%%%%%%%%%%%%%%%%%%%%%%%%%%%%%%%%%%%%%%%%%%%%%%%%%%%%
\subsection{Redundant bonds in  mixed-coordination Bethe networks}
\label{sec:redund-bonds-mixed}
As  discussed in recent work~\cite{DJTFMA99}, in order to find the
density $r$ of redundant links we need to integrate the density of
overconstrained links (as done in \Sec{sec:exact-calc-redund} for
Bethe Networks).  A link or bond is overconstrained whenever both its
end sites are outwards-rigid. A bond that connects a site of type
$\rho$ to a site of type $\nu$ is therefore overconstrained with
probability $T_\rho T_\nu$. If a fraction $X_{\nu \rho}$ of all links
connect sites of type $\nu$ to sites of type $\rho$, the average
density of overconstrained links is \hbox{$X_{11} T_1^2+ X_{12}
  T_1T_2+ X_{22} T_2^2$}. On the other hand, one can easily show that
\begin{equation} 
\left \{
\begin{array}{rcl}
X_{12} &=& \peff_1 p_{12} + \peff_2 p_{21}
\\ \\
X_{11} &=& \peff_1 p_{11}\\ \\
X_{22} &=& \peff_2 p_{22}.
\end{array}
\right .
\end{equation}
Using (\ref{eq:29}), the density of overconstrained bonds that has to
be integrated in order to obtain the density of redundant bonds is
then
\begin{eqnarray}
X_{11} T_1^2+ X_{12} T_1T_2+ X_{22} T_2^2 =
(\peff_1 T_1 + \peff_2 T_2 )^2 =
 T^2 .
\end{eqnarray}
Therefore,
\begin{equation}
  r = \int_0^b T^2 db,
\label{eq:9}
\end{equation}
which looks exactly the same as (\ref{eq:26}), but bear in mind that,
here, $T$ is given by (\ref{eq:15}), i.e~it is the average of $T_1$
and $T_2$ with weights $\peff_1$ and $\peff_2$. The simplicity of
(\ref{eq:9}) is a consequence of the random-bond hypothesis. While the
case of chemical order will not be considered here, we can anticipate
that, whenever the network has chemical order, the density of
overconstrained links no longer equals $T^2$.  In other words, for a
network with chemical order, the probability for a randomly chosen
link to be overconstrained is not simply the product of the a-priori
probabilities that its end sites be outwards-rigid (which is $T$ for
each site).

Now, since $x=bT$ (\Eqn{eq:15}), and, following the same steps as in
\Sec{sec:analyt-expr-r}, one has
\begin{eqnarray}
  r &=&2 \int_0^x T(x) dx - \left . x T \right |_0^x =
\nonumber \\ 
&= &
\peff_1 \left (2 \int_0^x T_1 dx - \left . x T_1 \right |_0^x \right )+
 \peff_2 \left (2 \int_0^x T_2 dx - \left . x T_2 \right |_0^x \right ).
\nonumber \\ 
\end{eqnarray}
Using (\ref{eq:12}) and (\ref{eq:27}), this can be written as
\begin{eqnarray}
r &=& 
\frac{\peff_1}{z_1} \sum_{j=g_1+1}^{z_1} (j-2g_1) P_j^{z_1}(x)
+
\frac{\peff_2}{z_2} \sum_{j=g_2+1}^{z_2} (j-2g_2) P_j^{z_2}(x),
\nonumber \\
%&=& x T - 2 \left ( \frac{\peff_1 g_1}{z_1} \sum_{j=g_1+1}^{z_1}  P^{z_1}_j(x)  
%+ \frac{\peff_2 g_2}{z_2} \sum_{j=g_2+1}^{z_2}  P^{z_2}_j(x) \right )
%\nonumber \\
\label{eq:rfinal}
\end{eqnarray}
which equals the average of the redundant-link density $r_\rho$ for
each homogeneous system (See (\ref{eq:27})), with weights
$\peff_\rho$.
\subsection{Continuity of $r$}
\label{sec:continuity-r}
Whenever there are multiple stable solutions for $T(b,p_1)$, the
ambiguity is raised by requiring that $r$ be maximal and a continuous
function of the parameter that drives the transition (either $b$ or
$p_1$). This is done by comparing the values of $r$ (\Eqn{eq:rfinal})
for the various branches with the help of either (\ref{eq:16}) or
(\ref{eq:17}).

The resulting final rigidity probabilities $T$ are shown in
\Fig{fig:vsgamma} for fixed $p_1$ and varying $b$, and in
\Fig{fig:vsp1} for fixed $b$ and varying $p_1$.  In both cases one
finds instances where two RP transitions exist, for certain ranges of
parameters. These ranges turn out to be rather narrow for the cases
displayed in \Figs{fig:vsgamma} and \ref{fig:vsp1}, which correspond
to $g_1=g_2=2$, $z_1=4$, and $z_2=20$. In fact it is easier to find
double transitions when there is both coordination contrast and dof
contrast, but we chose to analyze a case with $g_1=g_2$ and only
coordination contrast, because this is similar to what happens in
chalcogenide glass, where all atoms have the same number of dof (six
in three dimensions).

The connectivity range in which two $p_1$-driven RP transitions occur
is indeed so narrow that it is not observed in \Fig{fig:vsp1}. In
order to more clearly appreciate the double transitions, we have made
dedicated plots which show the double transitions in more
detail. Figures \ref{fig:2vsgamma} and \ref{fig:2vsp1} show zoomed-in
plots of the double transitions, respectively for the $b$-driven and
$p_1$-driven cases. The plots of $r$ (Figures \ref{fig:2vsgamma}c and
\ref{fig:2vsp1}c) show that, associated with double transitions, there
are two successive Maxwell loops in $r$.

Several observations can be done regarding the intermediate phase, or
first rigid phase, for this example: 1) The intermediate phase, as
already discussed in \Sec{sec:use-r-locate} is isostatic right at the
transition from the floppy phase.  2) The density $r$ or redundant
constraints is rather small in the intermediate phase, much smaller
than in the second rigid phase. A similar observation can be done
regarding the density $T^2$ of overconstrained links. 3) In the
intermediate phase $T_2\approx 0.9$ while $T_1\approx 0.2$ (See
Figures \ref{fig:2vsgamma}b and \ref{fig:2vsp1}b), so that the strong
component (sites with coordination $z_2=20$) rigidizes much more than
the weak component (sites with $z_1=4$).  At the second phase
transition, where the second rigid phase starts, the weak component
rigidizes abruptly, from $T_1 \approx 0.2$ to $T_1 \approx 0.8$, while
the jump in $T_2$ is there almost unnoticeable. Therefore, although
both components suffer discontinuities in $T$ at both transitions, in
practice one can say that the strong component $2$ rigidizes at the
first transition, while the weak component $1$ does so only at the
second transition. In the intermediate phase, the density of
overconstrained links $T^2$ is rather small of the order of $0.1$,
while it becomes of around $0.6$ at the second transition. Therefore,
the intermediate phase obtained in this model is rigid and, although
not isostatic, it is practically stress-free compared to the second
rigid phase.
\section{Discussion}
\label{sec:discussion}
We have presented an analytical study of Rigidity Percolation (RP) on
randomly bond-diluted (link density $b$) Bethe networks made from a
binary mixture of sites, respectively with fractions $p_1$ and $p_2$,
coordination numbers $z_1$ and $z_2$, and having $g_1$ and $g_2$
degrees of freedom (dof) at each site. Bond dilution can be thought of
as representing temperature effects in an approximate way. While
chemical order, i.e.~preferential bonding or dilution, can also be
studied with the methods discussed here, in this work only random
bonding (no chemical order) and homogeneous dilution was analyzed. Our
main result is that, for certain parameter combinations, two RP phase
transitions, either $b$-driven or $p_1$-driven, are found. A previous
numerical study by Thorpe and coworkers on a related problem~(in
\cite{TDRTA99}), found no evidence of two phase transitions.

A key step in this study was the calculation of the density $r$ of
redundant links, which results from integrating the density $T^2$ of
overconstrained links, as given by \Eqn{eq:26}. In previous work on
homogeneous Bethe networks~\cite{DJTFMA99}, this integral was done
numerically, while $T$ was found by iterating (\ref{eq:1}). In
contrast to this, we have shown that \Eqn{eq:26} can be explicitely
integrated, so that $r$ can be calculated analytically, resulting in
\Eqn{eq:27} for homogeneous Bethe networks, and (\ref{eq:rfinal}) for
mixed Bethe networks. For homogeneous networks, having an analytical
expression for $r$ allows one to determine the transition point
exactly (\Eqn{eq:bethe0}) for a few simple examples, and our results
are in good agreement with previous numerical work~\cite{DJTFMA99}. An
analysis of the transition-point equation in the form of \Eqn{eq:2}
furthermore allows one to conclude that, right at the transition, the
rigid cluster is isostatic, i.e. satisfies constraint balance.

On mixed Bethe networks it is found that, for certain combinations of
parameters, two RP phase transitions instead of one, are found. The
resulting intermediate phase is usually rather narrow. It is easier to
obtain double transitions for networks with $g_1\neq g_2$ and $z_1\neq
z_2$, i.e.~both coordination contrast and dof contrast. However, in
this analysis we concentrated on showing that a model binary glass,
with $g_1=g_2=2$ and $z_1 \neq z_2$, shows two transitions. Fixing
$z_1=4$, which is the minimum required to obtain a rigid structure
when $g=2$, we find that $z_2=16$ is the minimum required coordination
of the ``strong'' component, in order to obtain two RP
transitions. However, in this case the intermediate phase is extremely
narrow, so we chose to analyze $z_2=20$ instead, since its
intermediate phase is stronger. \Figs{fig:vsgamma} and \ref{fig:vsp1}
show, respectively, the $b$-driven and the $p_1$ driven transitions
for a broad range of parameters, while \Figs{fig:2vsgamma} and
\ref{fig:2vsp1} concentrate on those cases for which an intermediate
phase exists.

The intermediate phase is only isostatic at one of its endpoints, but
it can be argued that its self-stress is everywhere rather low,
compared to that of the second rigid phase. Furthermore, we have
observed that, approximately speaking, the strong component rigidizes
abruptly at the first transition ($T_2$ jumps form zero to a large
value) , while the weak one only does so at the second transition,
remaining only very weakly rigid ($T_1$ is small) in the intermediate
phase.

Our results, therefore, suggest an alternative mechanism for the
appearance of an intermediate phase in glass, which does not derive
from a self-organization principle but is a consequence of the
heterogeneity in the properties of the components forming the network.
Although the resulting intermediate phase is not isostatic, it can be
argued that its self-stress is very low compared to that of the second
rigid phase. Notice that available experimental results on glass point
to the existence of an intermediate phase with low stress but they
cannot determine whether this phase is rigorously isostatic or not.

Thorpe et al~\cite{TJCSIN00,TCJNIN01,CTSAR01}, on the other
hand, explain the intermediate phase as being a consequence of
self-stress minimization. Their proposed mechanism would give rise to
an isostatic intermediate phase, even on networks made from a single
component. We have noted that self-stress minimization does not
necessarily mean elastic-energy minimization, as the elastic modulus
of nearly isostatic structures is very low. Therefore, under strong
enough pressure, an isostatic structure would in fact maximize the
total elastic energy. This observation suggests a possible
experimental test of the proposed isostatic properties of the
intermediate phase.

Clearly, self-organization a la Thorpe et al, can be included in a
model as the one discussed here~\cite{BBLAAI05}, and it would be
interesting to see what results from the interplay of two different
mechanisms giving rise to multiple phase transitions. An important
ingredient that is still lacking in our present analysis is chemical
order, i.e~the fact that different atoms bond preferentially to atoms
of the same or a of a different species. The introduction of chemical
order is at present work in progress. Chemical order complicates the
treatment slightly, but appears feasible nevertheless.  

Numerical studies intended to corroborate the results presented in
this work are under way, and will be published elsewhere. Apart from
mixed Bethe networks, of course it is of interest to analyze
numerically three-dimensional models of chalcogenide glass, in order
to verify if the mechanism for intermediate phase described here is
still valid for more realistic cases.  

\acknowledgements 

The author acknowledges financial support from Conacyt under SNI
program.

%\bibliographystyle{unsrt}
%\bibliographystyle{abbrv}
%\bibliographystyle{plain}
%\bibliographystyle{apsrev}
%\bibliographystyle{apsrmp}
%\bibliography{GenRig,Glass,Percolation,Ripe,Moukarzel,Books}
\end{document}